\documentclass[reprint,
	amsmath,amssymb,
	aps,
	pra,
	floatfix,
]{revtex4-2}

\usepackage{graphicx}\usepackage{dcolumn}\usepackage{bm}\usepackage{hyperref}\usepackage{svg}    
	\usepackage{braket}
	\usepackage{mathtools}
	\usepackage{color}
	\usepackage{orcidlink}

\hypersetup{
 colorlinks = true,
 linkcolor = {blue},
 citecolor = {blue},
 urlcolor = {blue},
}

\usepackage[whole]{bxcjkjatype}
\usepackage[normalem]{ulem}
\usepackage{soul}

\begin{document}

\title{Quantum synchronization of qubits via dynamical Casimir effect}

\author{Haruki Mitarai\orcidlink{0000-0002-8374-3648}}
 \email{hmitarai@tkl.iis.u-tokyo.ac.jp}\author{Yoshihiko Hasegawa\orcidlink{0000-0001-7109-0611}}\email{hasegawa@biom.t.u-tokyo.ac.jp}
\affiliation{Graduate School of Information Science and Technology, \\
	The University of Tokyo, Tokyo 113-8656, Japan
	}

\date{\today}

\begin{abstract}
	In this paper, we study the synchronization of qubits induced by the dynamical Casimir effect
	in an atom-cavity quantum electrodynamics system. 
	Our investigation revolves around a pragmatic configuration of a quantum system, 
	where two superconducting qubits are coupled to a shared coplanar waveguide resonator 
	terminated at one end by a superconducting quantum interference device. 
The theoretical analyses of the system dynamics reveal sufficient conditions for ensuring synchronization which are anticipated to be accomplished by photon generation in the resonator. 
By numerically analyzing the time evolution of the system, 
	we confirm that the conditions are satisfied by photon generation via the dynamical Casimir effect, 
	resulting in qubit synchronization. Notably, we unveil a remarkable feature that is unique to synchronization induced by the dynamical Casimir effect: 
	the differences in the initial states of qubits and the differences in the coupling strengths of qubits to an electromagnetic field affect the synchronization independently without overlap between these factors.
\end{abstract}

\maketitle

\section{Introduction}
	Modern quantum field theory posits that real particles are generated due to vacuum fluctuations. 
	In 1970, Moore predicted that photons are generated from the vacuum by an accelerated mirror~\cite{moore1970quantum}, 
	and this phenomenon is now known as the dynamical Casimir effect~(DCE)~\cite{Dodonov2020fifty}.
	Importantly, the observable manifestation of photon production through the DCE requires the movement of the mirror at relativistic speeds.
However, it is highly challenging to achieve such substantial speeds for a massive mirror~\cite{PhysRevLett.96.200402, Dodonov2020fifty}. 
	Therefore, researchers have redirected their endeavors toward alternative methods that circumvent this challenging requirement, 
	namely the manipulation of the field boundary conditions by effective motion, 
	which has become the mainstream approach for observing the DCE~\cite{Yablonovitch1989, Dezael_2010, wilson2011observation, DCE-in-Josephson, Faccio_2011, vezzoli2019optical}. 
	
	In the late 2000s and early 2010s, a pragmatic experimental configuration leveraging superconducting microwave circuits came to the fore~\cite{PhysRevLett.103.147003, dce_microwave_circuits, PhysRevA.87.043804}. 
This watershed development signified a pivotal milestone in the empirical investigation of the DCE.
Notably, the first-ever observation of the DCE has been made using a coplanar waveguide terminated at one end by a superconducting quantum interference device (SQUID)~\cite{wilson2011observation}.
	This groundbreaking experimental result has motivated researchers to investigate potential applications of the DCE. Several studies have demonstrated the application of DCE in entanglement generation~\cite{PhysRevLett.124.140503, felicetti2014dynamical, PhysRevA.99.052328, DCE2022,Entangling_polaritons}
	and the Gaussian boson sampling~\cite{bosonsampling}. 
	These studies suggest that the DCE can serve as a valuable resource for quantum technologies. 

	Along with the DCE, the core element of the present study is the intriguing phenomenon of synchronization.
	Synchronization, characterized by an adjustment of rhythms or phases of self-sustained oscillators due to their weak interaction or external driving, 
	is omnipresent in a wide range of fields~\cite{Cambridge_2001}.
For example, in the field of neuroscience, the synchronous firing of neurons is crucial to the rapid pattern recognition of visual information for mammals~\cite{Visual_neuron}.
	Further, since the last decade, synchronization in quantum regimes has attracted a great deal of interest.
	Many studies have focused on the synchronization in quantum systems, 
	such as quantum van der Pol oscillators~\cite{Quantum_synchronization_of_two_Van_der_Pol_oscillators, Nissi, Walter_2014, Squeezing_Enhances, setoyama2024lie}, 
	two-level systems~\cite{Zhirov, Zhirov2, eneriz2019degree, bath2021, star-type, huan2020synchronization, composite-two-qubit, Classical-driving}, 
	noise-induced synchronization \cite{Noise-Induced,setoyama2023lie}, 
and others~\cite{Spin-1_Atoms,Measure_synchronization, Quantum_Synchronization_and_Entanglement, Measurement-Induced, Topological}.
Moreover, several studies have explored the potential applications of quantum synchronization, 
	e.g., in a probing scheme and quantum communication~\cite{10.1063/5.0061478, PhysRevA.94.052121, Liu:19}, 
highlighting the potential of quantum synchronization in the significant advancement of the frontiers of quantum science.

	In this paper, we investigate quantum synchronization of two qubits induced by the DCE.
 	The state-of-the-art superconducting qubits, 
	constructed using a coplanar waveguide resonator and transmons, are cooled to 10--30~mK, where very few thermal photons exist in the resonator~\cite{nakamura1999coherent,blais2004cavity, demonstration,clarke2008superconducting, Charge-insensitive,IBM_qubit}.
	Thus, quantum theoretical effects such as the DCE, 
	which are not considered in classical systems, would have a great impact on the superconducting qubit.
	Note that the interplay between the qubits and the DCE is not exclusively confined to our study. Several researchers have explored the DCE in the atom-cavity quantum electrodynamics (QED) systems~\cite{felicetti2014dynamical, Entangling_polaritons, vyatkin2023resonant, DCE_QIP,two-atom, Veloso_2015, DODONOV2020126837, PhysRevA.98.022520}.
For example, the DCE in the two-atom setup has been studied in Ref.~\cite{two-atom}. 
	The authors have presented that photons are produced even when the cavity interacts with the two artificial atoms, 
resulting in enhanced atomic excitation probabilities. 
Additionally, Ref.~\cite{DCE_QIP} has pointed out the strong impact of photon emission by the DCE on quantum information processing.
Thus, the following intriguing and novel prospect can be explored based on the profoundness of the DCE on quantum systems: 
the potential application of the DCE as an external drive for qubit synchronization.

	Here, we assume a realistic construction of a quantum system,
	where two superconducting qubits are coupled to the shared coplanar waveguide resonator, 
	that is terminated at one end by a SQUID.
By modulating the external flux threaded through the SQUID, 
	the boundary condition of the resonator changes, 
	and then DCE photons are generated in the resonator~\cite{PhysRevLett.103.147003, dce_microwave_circuits}.
However, as the computational task becomes more challenging with the increasing number of generated photons in the resonator, 
	we turn off the oscillation of the boundary condition after a set period.
The theoretical analyses of the system dynamics reveal sufficient conditions for achieving in-phase synchronization under resonance conditions, 
	indicating the possibility of inducing synchronization via photon generation.
	Our main findings center on the demonstration of qubit synchronization in scenarios 
where one qubit exhibits a divergent initial state from the other, 
	where each qubit interacts with a resonator through distinct coupling strengths, 
	and where one of the qubit transition frequencies is detuned.
	Remarkably, we reveal a characteristic that is unique to DCE-induced qubit synchronization: 
the differences in the initial states and the coupling strengths independently affect qubit synchronization, 
	without any overlap between these factors.

\section{Model} \label{sec:model}
	\begin{figure}[t]
		\vspace{0.32cm}
		\begin{minipage}[t]{1\hsize}
			\centering
			\includegraphics[width = 5.5cm]{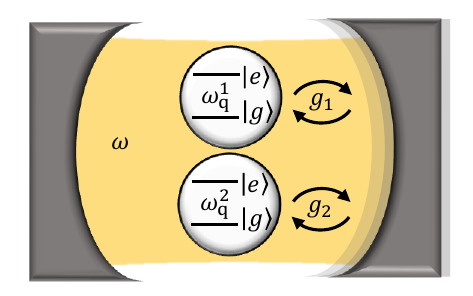} \\
			\vspace{-3.7cm}\hspace{-4.3cm}(a) \\
		\end{minipage}\\
		\begin{minipage}[t]{1\hsize}
			\centering
			\includegraphics[width = 8.0cm]{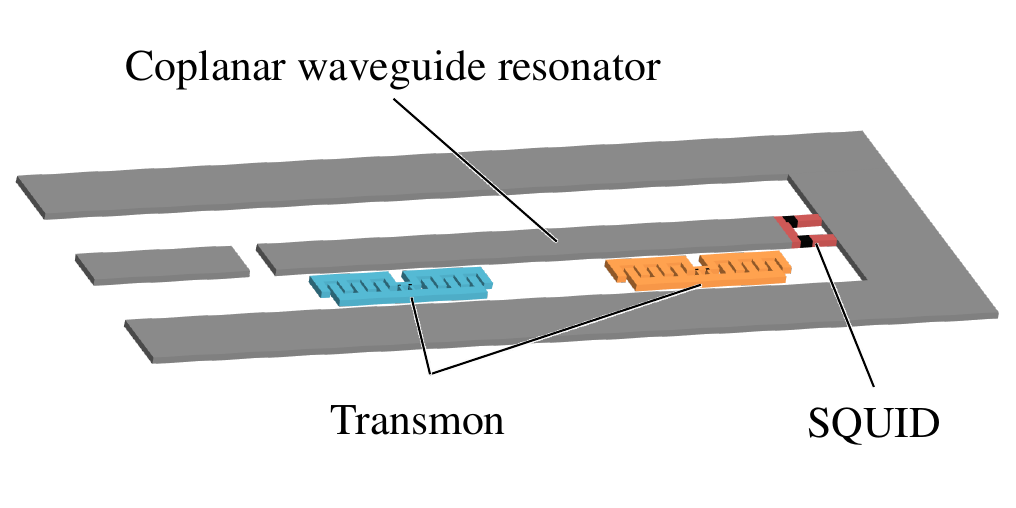} \\
			\vspace{-3.5cm}\hspace{-7.5cm}(b)
		\end{minipage}
		\caption{(a)~Conceptual illustration of the model comprising an electromagnetic field in the single-mode cavity and two qubits.
			The two qubits are coupled to the electromagnetic field in the cavity with coupling strengths $g_{1}$ and $g_2$, respectively.
			Here, $\omega$ is the cavity resonance frequency, 
			and $\omega_\mathrm{q}^{1,2}$ are the atomic transition frequencies. 
			(b)~Simplified schematic layout of implementation of the model.
			The coplanar waveguide resonator is coupled to two transmons.
			The boundary condition of the resonator is changed by modulating the external flux threading the SQUID terminating one end of the resonator.
		}
		\label{fig:model}
	\end{figure}

	In this section, we present a model comprising two qubits coupled to a single-mode cavity shown in Fig.~\ref{fig:model}(a).
	The oscillating boundary condition of the cavity results in photon generation by virtue of the DCE.
	Figure~\ref{fig:model}(b) shows the schematic of our system on a realistic superconducting circuit.
The coplanar waveguide resonator acts as an electromagnetic field cavity, and the transmons act as two-level systems~\cite{Charge-insensitive}.
The boundary condition of the resonator is changed by modulating the external flux through the SQUID terminating one end of the resonator.

	The Hamiltonian describing the system illustrated in Fig.~\ref{fig:model}(b) is expressed as follows~\cite{felicetti2014dynamical}: 
	\begin{equation}
		\label{eq:H}
		H \equiv H_{\mathrm{TC}} + \alpha\left(t\right) \left( a^{\dagger}+a \right)^2,
	\end{equation}
	with Tavis-Cummings Hamiltonian~\cite{TCM}
	\begin{equation}
		\label{eq:H_TC}
		H_{\mathrm{TC}} \equiv \omega a^{\dagger} a + \sum_{\mu=1}^2{\left[\frac{\omega_{\mathrm{q}}^{\mu}}{2} \sigma_{\mathrm{z}}^{\mu} + g_{\mu} \left( \sigma_{\mathrm{-}}^\mu a^{\dagger} + \sigma_{\mathrm{+}}^\mu a \right)\right]}.
	\end{equation}
	Here, $a^\dagger~(a)$ is a creation~(annihilation) operator of the electromagnetic field in the cavity with the resonance frequency $\omega$, 
	and $\sigma_\mathrm{z,\pm}^\mu$ are the Pauli, raising, and lowering operators associated with the qubit~$\mu$ with an atomic transition frequency $\omega_\mathrm{q}^\mu$.
	The coupling strengths between the electromagnetic field and qubit~$\mu$ are denoted as $g_\mu$.
	The oscillation of the cavity boundary condition caused by the oscillation of the Josephson energy of the SQUID is denoted by 
	$\alpha \left(t\right) = \alpha_0 \cos\left(\omega_\mathrm{d}t\right) \Theta\left( \tau - t \right)$, 
	where $\Theta\left( t \right)$ is the Heaviside step function, $\omega_\mathrm{d}$ is the driving frequency, 
	and $\tau$ is the predetermined time of turning off the oscillation.
	Owing to the extremely rapid increase in the mean photon number, we turn off the oscillation of $\alpha\left(t\right)$ using $\Theta\left( \tau - t \right)$ 
	to calculate the time evolution of the system with finite computational resources. 
	Throughout the present paper, $\tau = 500$ is adopted.
	Additionally, we use $\hbar=1$ and treat the cavity as a single-mode resonator. 

	To enhance clarity, 
	we move into the frame rotating at the driving frequency $\omega_\mathrm{d}$.
	Let $\Ket{\psi_{\mathrm{I}}\left(t\right)} \equiv \mathrm{e}^{iH_0t} \Ket{\psi\left(t\right)}$ be the state vector in the rotating frame
	with $H_0 \equiv \omega_\mathrm{d} a^{\dagger} a + \frac{\omega_{\mathrm{d}}}{2} \sum_{\mu=1}^2{\sigma_{\mathrm{z}}^{\mu}}$
	and $\Ket{\psi\left(t\right)}$ being the state vector in the Schr\"{o}dinger picture.
	The time evolution of the system is described by 
	the Schr\"{o}dinger equation in the frame rotating at $\omega_\mathrm{d}$; 
	$\dot{\Ket{\psi_{\mathrm{I}}}} = -i H_{\mathrm{I}} \Ket{\psi_{\mathrm{I}}}$, where 
	\begin{align}
		\label{eq:HI}
		H_{\mathrm{I}} \equiv
		\left(\omega - \frac{\omega_{\mathrm{d}}}{2}\right)a^{\dagger}a
		&+\sum_{\mu=1}^2{ \left[\frac{\omega_{\mathrm{q}}^{\mu} - \frac{\omega_{\mathrm{d}}}{2}}{2}
		+ g_{\mu} \left( \sigma_{\mathrm{-}}^\mu a^{\dagger} + \sigma_{\mathrm{+}}^\mu a \right)\right]} \nonumber\\
		&+ \frac{\alpha_0}{2} \Theta\left( \tau - t \right) \left( {a^{\dagger}}^2 + a^2\right).
	\end{align}
	The fast oscillating terms are neglected in Eq.~\eqref{eq:HI} by means of the rotating-wave approximation (RWA).
Further, we assume the initial state of the system is prepared as 
	\begin{align}
		\label{eq:initial_state}
		\Ket{\psi(0)} &= \left(\cos\theta_1 \Ket{g}_1 + \sin\theta_1 \Ket{e}_1 \right) \nonumber\\ 
		&\quad\otimes \left(\cos\theta_2 \Ket{g}_2 + \sin\theta_2 \Ket{e}_2 \right) \otimes \Ket{0},
	\end{align}
where $\Ket{g}_\mu$ ($\Ket{e}_\mu$) represents the ground (exited) state of the $\mu$th qubit, 
	and $\Ket{0}$ represents the vacuum state of the cavity.
	
	Our research focuses on the synchronization of $\braket{\sigma_\mathrm{z}^{1}}$ and $\braket{\sigma_\mathrm{z}^{2}}$, 
	where $\braket{O}$ stands for the expectation value of $O$, 
	i.e., $\braket{O} \equiv \braket{\psi\left(t\right)|O|\psi\left(t\right)}$.
	To quantify the degree of synchronization between $\braket{\sigma_\mathrm{z}^{1}}$ and $\braket{\sigma_\mathrm{z}^{2}}$, 
	we use the Pearson correlation coefficient defined as~\cite{Galve2017} 
	\begin{align}
		\label{eq:PCC}
		C_{\Delta t}\left(t\right)&\equiv\frac{\int_{t}^{t+\Delta t}{\left(\Braket{\sigma_{\mathrm{z}}^{1}\left(t^{\prime}\right)}-\bar{\sigma}_{\mathrm{z}}^{1}\right)\left(\Braket{\sigma_{\mathrm{z}}^{2}\left(t^{\prime}\right)}-\bar{\sigma}_{\mathrm{z}}^{2}\right)\mathrm{d}t^{\prime}}}{\sqrt{\prod_{\mu=1}^{2}\int_{t}^{t+\Delta t}{\left(\Braket{\sigma_{\mathrm{z}}^{\mu}\left(t^{\prime}\right)}-\bar{\sigma}_{\mathrm{z}}^{\mu}\right)^{2}}\mathrm{d}t^{\prime}}},
	\end{align}
	with
	\begin{equation}
		\label{eq:bar_sigma_z}
		\bar{\sigma}_{\mathrm{z}}^\mu \equiv \frac{1}{\Delta t} \int_{t}^{t + \Delta t}{\Braket{\sigma_{\mathrm{z}}^\mu \left(t^\prime\right)}\mathrm{d}t^\prime},
	\end{equation}
	and $\Delta t$ being a time window.
 
\section{Theoretical results} \label{sec:analitical}
	\subsection{Balanced case} \label{sec:balanced}
		In the subsequent numerical analysis in Sec.~\ref{sec:numerical_results}, 
		the boundary oscillation is turned off after a predetermined period $\tau$, 
		as described in Sec.~{\ref{sec:model}}.
		Prior to presenting the numerical results, 
		we analyze the behaviors of the two-atom TCM with and without DCE driving, namely the behaviors in $t \le \tau$ and $t > \tau$.
		Thereafter, we reveal the conditions that yield the synchronization of $\braket{\sigma_\mathrm{z}^{1}}$ and $\braket{\sigma_\mathrm{z}^{2}}$.

		First, we expand $\Ket{\psi_{\mathrm{I}}}$ as follows:
		\begin{align}
			\label{eq:Phi_I_expand2}
			\Ket{\psi_{\mathrm{I}}} = &\sum_{m=0}^\infty \left[ \mathcal{A}_m\left(t\right)  \Ket{g}_1 \Ket{g}_2 \Ket{m} \right. 
			+\mathcal{B}_m\left(t\right)  \Ket{g}_1 \Ket{e}_2 \Ket{m} \nonumber\\
			&+\mathcal{C}_m\left(t\right)  \Ket{e}_1 \Ket{g}_2 \Ket{m} 
			\left. +\mathcal{D}_m\left(t\right)  \Ket{e}_1 \Ket{e}_2 \Ket{m} \right], 
		\end{align}
		where $\ket{m}$ represent the cavity Fock states with $m$ photons.
		The differential equations for $\mathcal{A}_m, \mathcal{B}_m, \mathcal{C}_m$, and $\mathcal{D}_m$ are derived from 
		the Schr\"{o}dinger equation, i.e.,  
		$\dot{\Ket{\psi_{\mathrm{I}}}} = -i H_{\mathrm{I}} \Ket{\psi_{\mathrm{I}}}$.
		
		We begin by considering the balanced scenario $g_1 = g_2 = g$. 
		Additionally, we assume that the two qubits and the cavity are in resonance, 
		where $\omega_{\mathrm{q}}^1 = \omega_{\mathrm{q}}^2 = \omega$.
		Regarding $t > \tau$, the Schr\"{o}dinger equation in the rotating frame can be solved analytically. 
		For $m \ge 2$, $\mathcal{A}_m, \mathcal{B}_m, \mathcal{C}_m$, and $\mathcal{D}_m$ become~\cite{two-atom}
		\begin{align}
			\label{eq:abcd_1a}
			&\mathcal{A}_m\left(t\right) =
			\frac{C_{4,m}\left(1-m\right) + C_{3,m} m \mathrm{e}^{i g K_m t} + C_{2,m} m \mathrm{e}^{-i g K_m t}}{\sqrt{m}\sqrt{m-1}},\\
			\label{eq:abcd_1b}
			&\mathcal{B}_{m-1}\left(t\right) = -C_{1,m} \nonumber\\
			&\qquad\qquad + \frac{C_{3,m} K_m \mathrm{e}^{i g K_m t} - C_{2,m} K_m\mathrm{e}^{-i g K_m t}}{2\sqrt{m-1}},\\
			\label{eq:abcd_1c}
			&\mathcal{C}_{m-1}\left(t\right) = \mathcal{B}_{m-1} + 2C_{1, m},\\
			\label{eq:abcd_1d}
			&\mathcal{D}_{m-2}\left(t\right) = C_{4,m} + C_{3,m} \mathrm{e}^{i g K_m t} + C_{2,m} \mathrm{e}^{-i g K_m t},
		\end{align}
		with $K_m \equiv \sqrt{2 \left(2m - 1\right)}$.
		For $m = 1$ and $t > \tau$, 
		\begin{align}
			\label{eq:abcd_2a}
			\mathcal{A}_1\left(t\right) &= \sqrt{2} \left(C_{2,1}  \mathrm{e}^{- i \sqrt{2}g t} - C_{3,1}  \mathrm{e}^{i \sqrt{2}g t} \right),\\
			\label{eq:abcd_2b}
			\mathcal{B}_{0}\left(t\right) &= C_{2,1}  \mathrm{e}^{- i \sqrt{2}g t} + C_{3,1}  \mathrm{e}^{i \sqrt{2}g t} - C_{1,1} , \\
			\label{eq:abcd_2c}
			\mathcal{C}_{0}\left(t\right) &= \mathcal{B}_{1} + 2C_{1, 1},
		\end{align}
		where $C_{j,m}$ are complex constant coefficients.
		Regarding $t \le \tau$, where the boundary condition oscillates, $C_{j,m}$ become time-dependent.
		
		Now, to theoretically explore the in-phase synchronization, we introduce $\braket{\Delta \sigma_\mathrm{z}} \equiv \braket{\sigma_\mathrm{z}^{1}}-\braket{\sigma_\mathrm{z}^{2}}$. 
		By defining $\Braket{\Delta \sigma_{\mathrm{z},m}} \equiv - \left| \mathcal{B}_m \right|^2 + \left| \mathcal{C}_m \right|^2$, 
		we can express $\braket{\Delta \sigma_\mathrm{z}}$ in terms of $\Braket{\Delta \sigma_{\mathrm{z},m}}$ as follows:
		\begin{equation}
			\label{eq:pauli_z_diff}
			\braket{\Delta \sigma_\mathrm{z}\left(t\right)} = \sum_{m=0}^{\infty} \Braket{\Delta \sigma_{\mathrm{z},m}\left(t\right)}.
		\end{equation}
		Here, we can consider two sufficient conditions for  $C_{j,m}\left(\tau\right)$ to ensure $\braket{\Delta \sigma_\mathrm{z}\left(t > \tau\right)}=0$. 
		First, if $C_{1,m} = 0$ hold at $t = \tau$, then $\mathcal{B}_m\left(t>\tau\right) = \mathcal{C}_m\left(t>\tau\right)$, 
		which immediately leads to vanishing $\braket{\Delta \sigma_\mathrm{z}\left(t > \tau\right)}$. 
		Second, the satisfaction of the two conditions,
		\begin{equation}
			\label{eq:condition_A2}
			\left\{ \,
			\begin{aligned}
				&\left|C_{2,m}\right| - \left|C_{3,m}\right| = 0  \\
				&\arg\left[C_{2,m} + C_{3,m} \right] - \arg \left[C_{1,m}\right] = \pi\left(k+\frac{1}{2}\right) \quad \left( k \in \mathbb{Z} \right) 
			\end{aligned}
			\right.
		\end{equation}
		is also a sufficient condition for ensuring $\braket{\Delta \sigma_\mathrm{z}\left(t > \tau\right)} = 0$.
		Here, $\arg\left[\cdot\right]$ represents the argument of the complex.
		As it is evident from Eq.~\eqref{eq:HI}, 
$C_{j,m}\left(\tau\right)$ depend on $\alpha_0$ and $\tau$.
		Hence, 
		if the cessation of boundary oscillation coincides with the moment 
		when $C_{1,m} = 0$ or Eq.~\eqref{eq:condition_A2} is fulfilled, 
		it is predicted that $\braket{\sigma_\mathrm{z}^{1}}$ and $\braket{\sigma_\mathrm{z}^{2}}$ will subsequently synchronize. 
Sec.~\ref{sec:result1} presents the numerical results of this scenario.

		At this point, it is worth highlighting a significant difference between quantum and classical systems. 
		In classical systems, oscillators synchronize after a certain period if they share a common support and possess similar characteristics~\cite{Cambridge_2001}. 
		The two-atom TCM under consideration, where the qubits share the cavity, exhibits a resemblance to this classical scenario.
		However, assuming $\alpha\left(t\right) = 0$ and $\exists m \ge 1, C_{1,m}\left(t=0\right) \ne 0$, 
		we can immediately deduce that the in-phase synchronization of $\braket{\sigma_\mathrm{z}^{1}}$ and $\braket{\sigma_\mathrm{z}^{2}}$ never occurs in the balanced case 
		since $C_{1,m}$ remains constant as long as $\alpha\left(t\right) = 0$. 
		In fact, $\left|C_{1,m}\right|^2$ represents the probability of detecting the dark state ${\left(\Ket{g}_1 \Ket{e}_2 \Ket{m} - \Ket{e}_1 \Ket{g}_2 \Ket{m}\right)}/{\sqrt{2}}$, 
		that is decoupled from the cavity field~\cite{alzetta1976experimental, Interfering_pathways, zanner2022coherent}.
		Therefore, we should keep in mind that the symmetrical configuration, i.e., $g_1 = g_2$ in the present context, does not necessarily guarantee synchronization within the quantum systems.

	\subsection{Unbalanced case} \label{sec:unbalanced}
		Next, we consider the unbalanced scenario where $g_1 \ne g_2$. 
		Here we assume that the qubits and the cavity are in resonance: $\omega_{\mathrm{q}}^1 = \omega_{\mathrm{q}}^2 = \omega$.
		Regarding $m \ge 2$ and $t > \tau$, 
		$\mathcal{A}_m, \mathcal{B}_m, \mathcal{C}_m$, and $\mathcal{D}_m$ are given by~\cite{two-atom}
		\begin{align}
			\label{eq:abcd_3a}
			\mathcal{A}&_m\left(t\right) = \sum_{\mathcal{S}=\pm1} \sum_{\mathcal{U}=\pm1} \nonumber\\
			&\quad\quad\quad\frac{C_{\mathcal{S},\mathcal{U}, m}\, {\mathrm{e}}^{i  \,\mathcal{U}\frac{\sqrt{2}M_{\mathcal{S}m} t}{2}} \left({g_{1}}^2+{g_{2}}^2 + \mathcal{S}L_{m}\right)}{4 g_{1} g_{2} \sqrt{m} \sqrt{m-1}}, \\
			\label{eq:abcd_3b}
			\mathcal{B}&_{m-1}\left(t\right) = \sum_{\mathcal{S}=\pm1} \sum_{\mathcal{U}=\pm1} \nonumber\\
			- &\mathcal{S}\mathcal{U}\frac{\sqrt{2} C_{\mathcal{S},\mathcal{U}, m} M_{\mathcal{S}m}\, {\mathrm{e}}^{i\,\mathcal{U}\frac{\sqrt{2} M_{\mathcal{S}m} t}{2}} \left(L_m + \mathcal{S}\left(G^2 - 4g_1^2 m\right)\right)}{8g_{1} m \left({g_{1}}^2-{g_{2}}^2\right) \sqrt{m-1}}, \\
			\label{eq:abcd_3c}
			\mathcal{C}&_{m-1}\left(t\right) = \sum_{\mathcal{S}=\pm1} \sum_{\mathcal{U}=\pm1} \nonumber\\
			&\mathcal{S}\mathcal{U}\frac{\sqrt{2} C_{\mathcal{S},\mathcal{U}, m} M_{\mathcal{S}m}\, {\mathrm{e}}^{i\,\mathcal{U}\frac{\sqrt{2} M_{\mathcal{S}m} t}{2}} \left(L_m + \mathcal{S}\left(G^2 - 4g_2^2 m\right)\right)}{8g_{2} m \left({g_{1}}^2-{g_{2}}^2\right) \sqrt{m-1}}, \\
			\label{eq:abcd_3d}
			\mathcal{D}&_{m-2}\left(t\right) = \sum_{\mathcal{S}=\pm1} \sum_{\mathcal{U}=\pm1}  C_{\mathcal{S},\mathcal{U}, m} \, {\mathrm{e}}^{i\,\mathcal{U}\frac{\sqrt{2} M_{\mathcal{S}m} t}{2}}, 
		\end{align}
		where,
		\begin{align*}
L_m \equiv &\sqrt{G^4 + 16 g_1^2 g_2^2 m \left( m-1 \right)}, \\
			M_{\pm m} \equiv &\sqrt{G^2\left(2m-1\right) \pm L_m}, \\
			G \equiv &\sqrt{g_1^2 + g_2^2}\,,
		\end{align*}
		and $C_{\mathcal{S},\mathcal{U}, m}$ are complex constant coefficients.
		Regarding $m = 1$ and $t > \tau$, we obtain, 
		\begin{align}
			\label{eq:abcd_4a}
			\mathcal{A}_1\left(t\right) &= G \left(C_{2}  \mathrm{e}^{- i G t} - C_{3}  \mathrm{e}^{i G t} \right),\\
			\label{eq:abcd_4b}
			\mathcal{B}_{0}\left(t\right) &= g_2 \left(C_{2}  \mathrm{e}^{- i G t} + C_{3}  \mathrm{e}^{i G t}\right) - C_{1}g_1,\\
			\label{eq:abcd_4c}
			\mathcal{C}_{0}\left(t\right) &=  g_1\left( C_{2}  \mathrm{e}^{- i G t} + C_{3}  \mathrm{e}^{i G t}\right) + C_{1}g_2,
		\end{align}
		where $C_j$ are complex constant coefficients. 
		For $t \le \tau$, $C_{\mathcal{S},\mathcal{U},m}\left(t\right)$ and $C_{j}\left(t\right)$ become time-dependent; 
		consequently, $C_{\mathcal{S},\mathcal{U},m}\left(\tau\right)$ and $C_{j}\left(\tau\right)$ depend on $\alpha_0$ and $\tau$ as the balanced case.
		Assuming $\left|g_1-g_2 \right| \ll \left|g_1\right|,\,\left|g_2\right|$, we obtain 
		\begin{align}
			&\frac{ L_m + \left(G^2-4g_1^{2}m\right)}{g_1\left(g_1^{2} - g_2^{2}\right)} \sim -\frac{ L_m + \left(G^2-4g_2^{2}m\right)}{g_2\left(g_1^{2} - g_2^{2}\right)}, \nonumber\\
			&\frac{ L_m - \left(G^2-4g_1^{2}m\right)}{g_1\left(g_1^{2} - g_2^{2}\right)} \sim \frac{ L_m - \left(G^2-4g_2^{2}m\right)}{g_2\left(g_1^{2} - g_2^{2}\right)}. \nonumber
		\end{align}
		Accordingly, we obtain two sufficient conditions for $C_{\mathcal{S}, \mathcal{U}, m}$, 
		where $\Braket{\Delta\sigma_{\mathrm{z},m}\left(t\right)} \sim 0$ is permanently maintained for $t > \tau$: 
		first, $C_{-1, \mathcal{U}, m} = 0$, and second, 
		\begin{equation}
			\label{eq:condition_B2}
			\left\{ \,
			\begin{aligned}
				\left|C_{-1, +1, m}\right| - \left|C_{-1, -1, m}\right| = &\; 0\\
				\left|C_{+1, +1, m}\right| - \left|C_{+1, -1, m}\right| = &\; 0 \\
				\arg\left[C_{-1, +1, m} + C_{-1, -1, m} \right] &- \arg \left[C_{+1, +1, m} + C_{+1, -1, m} \right] \\
				&= \pi\left(k+\frac{1}{2}\right) \quad \left( k \in \mathbb{Z} \right).
			\end{aligned}
			\right.
		\end{equation}
		The two conditions for $C_{j}\left(\tau\right)$ to ensure $\Braket{\Delta\sigma_{\mathrm{z},0}\left(t\right)} \sim 0$ are the same as the those for $C_{j, m}$: $C_1 = 0$ or 
		\begin{equation}
			\label{eq:condition_B3}
			\left\{ \,
			\begin{aligned}
				& \left|C_{2}\right| - \left|C_{3}\right| = 0 \\
				& \arg\left[C_{2} + C_{3} \right] - \arg \left[C_{1}\right] = \pi\left(k+\frac{1}{2}\right) \quad \left( k \in \mathbb{Z} \right).
			\end{aligned}
			\right.
		\end{equation}
		Hence, similar to the balanced scenario where $g_1 = g_2$, 
the DCE may be crucial to synchronizing $\sigma_\mathrm{z}^\mu$. 
Sec.~\ref{sec:result2} and Sec.~\ref{sec:result3} present the numerical results for the unbalanced scenario.

\section{Numerical results} \label{sec:numerical_results}
The Schr\"{o}dinger equation in the rotating frame is solved numerically to verify the occurrence of synchronization.
		The following four scenarios are considered:
		\begin{itemize}
			\item Balanced case with different initial states, where all the parameters are identical between the qubits, except for $\theta_\mu$ (Sec.~\ref{sec:result1}).
			\item Unbalanced case with the same initial states, where all the parameters are identical between qubits except for the coupling strength $g_\mu$ (Sec.~\ref{sec:result2}).
			\item Unbalanced case with different initial states, where the parameters of each qubit are identical to those of the other qubit, except for the $\theta_\mu$ and $g_\mu$ (Sec.~\ref{sec:result3}).
			\item Balanced case with the same initial states and frequency detuning, where all the parameters are identical between the qubits except for $\omega_{\mathrm{q}}^\mu$ (Sec.~\ref{sec:result4}).
		\end{itemize}
		For all scenarios, the calculations are performed for the Hamiltonian after the RWA, i.e., Eq.~\eqref{eq:HI}.

	\subsection{Balanced case with different initial states} \label{sec:result1}
		\begin{figure}[t]
			\begin{minipage}[t]{1\hsize}
				\centering
				\includegraphics[width = 8.6cm]{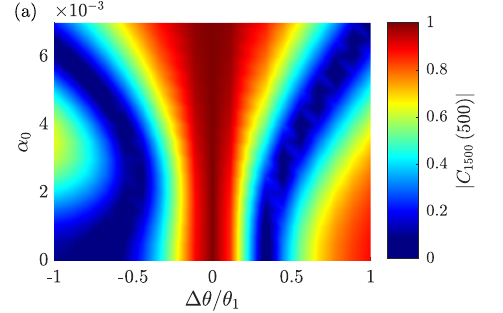}
					\label{fig:arnold_initialtTheta}
			\end{minipage}\\
			\begin{minipage}[t]{1\hsize}
				\centering
				\includegraphics[width = 8.6cm]{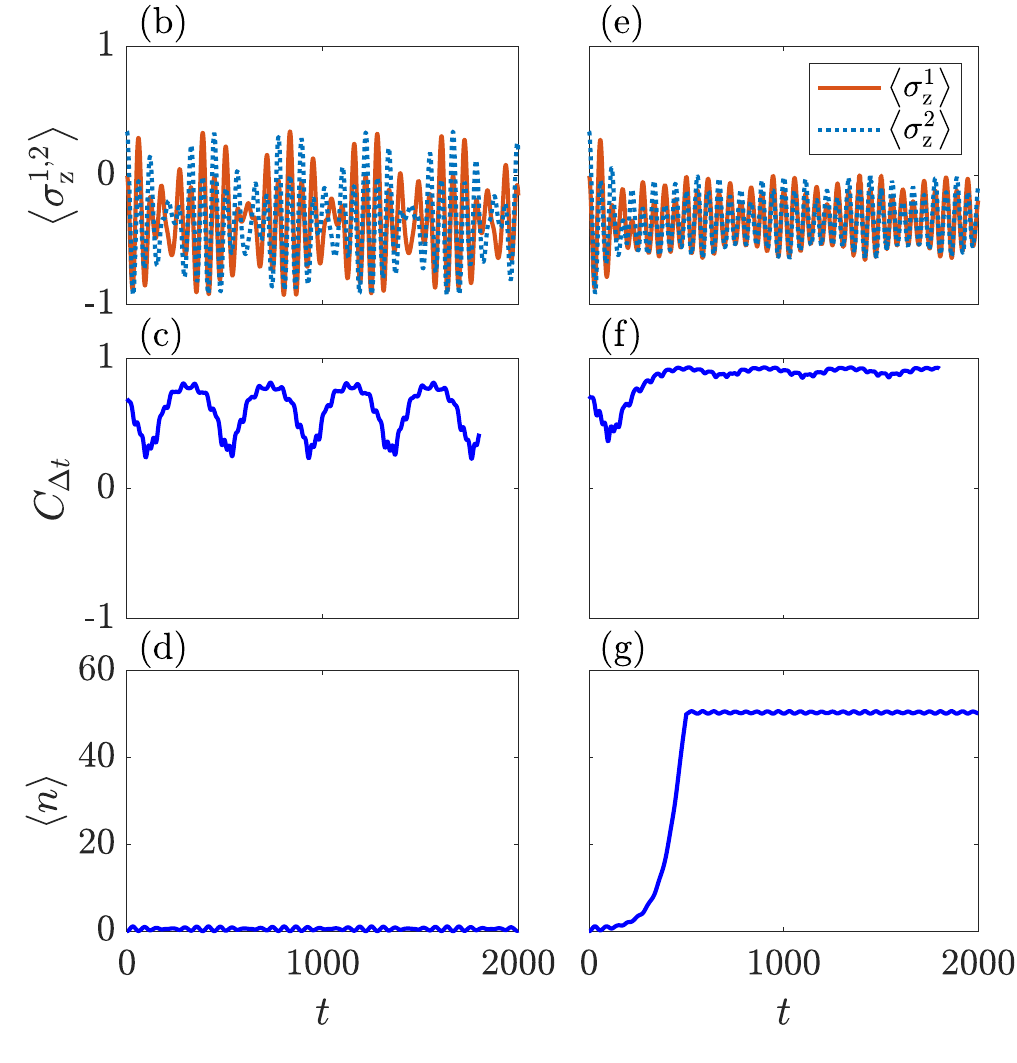}
			\end{minipage}
			\caption{Results of the balanced case with different initial states.
				The qubits are coupled to the electromagnetic field with the same coupling strengths $g_1 = g_2 = 0.04$.
				We set $\theta_1 = \pi/4$, $\omega = \omega_\mathrm{q}^1 = \omega_\mathrm{q}^2 = 1$, $\omega_\mathrm{d} = 2$, and $\tau = 500$ for (a)--(g).
				(a)~Pearson correlation coefficient $\left| C_{1500}\left(500\right) \right|$
				as a function of $\Delta \theta /\theta_1$ and $\alpha_0$, 
				where $ \Delta \theta \equiv \left(\theta_1 - \theta_2\right)$.
				(b) and (e) Time evolution of $\Braket{\sigma_\mathrm{z}^{1,2}}$.
				(c) and (f) $C_{\Delta t}\left(t\right)$ with $\Delta t = 200$.
				(d) and (g) Mean photon number $\Braket{n}\equiv\braket{a^\dagger a}$.
				We choose $\alpha_0=0$ for (b)--(d), and $\alpha_0=6 \times 10^{-3} $ for (e)--(g).
				We set $\theta_2 = 11\pi/36$ for (b)--(g).
				The abrupt change in the trend of the mean photon number at $t=500$ in (g) 
				is a consequence of the turning-off of the boundary oscillation.
			}
			\label{fig:result1}
		\end{figure}
		In this section, we show the occurrence of qubit synchronization 
		under conditions where all the parameters are identical between the qubits, except for the initial states. 
		We set $\theta_1 = \pi/4$, $g_1 = g_2 = 0.04$, $\omega = \omega_\mathrm{q}^1 = \omega_\mathrm{q}^2 = 1$, and $\omega_\mathrm{d} = 2$. 
		Fig.~\ref{fig:result1}(a)
		shows the Pearson correlation coefficient $\left| C_{1500}\left(500\right) \right|$
		as a function of $\Delta \theta /\theta_1$ and $\alpha_0$, where $ \Delta \theta \equiv \left(\theta_1 - \theta_2\right)$.
		At first glance, we can see that the range of $\Delta \theta$ 
with higher values of $\left| C_{1500}\left(500\right) \right|$ widens as $\alpha_0$ increases.
		The prerequisite for achieving synchronization is encapsulated in the condition $C_{1,m}=0$ or Eq.~\eqref{eq:condition_A2}, 
		as presented in Sec.~\ref{sec:balanced}. 
To clarify which condition is satisfied when synchronization occurs with $\theta_1 \ne \theta_2$, 
		we calculate $C_{j,1}\left(0\right)$ and $C_{j,1}\left(\tau\right)$. 
		For $\alpha_0=6 \times 10^{-3}$ with $\theta_1 = \pi/4$ and $\theta_2 = 11\pi/36$, 
		we obtain 
		\begin{align*}
			C_{1,1}\left(0\right)&\sim0.087\,\mathrm{e}^{-i\pi}, \\
			C_{2,1}\left(0\right)&\sim0.246\,\mathrm{e}^{0}, \\
			C_{3,1}\left(0\right)&\sim0.246\,\mathrm{e}^{0}, 
	   	\end{align*}
		and
		\begin{align*}
			 C_{1,1}\left(\tau\right)&\sim0.027\,\mathrm{e}^{-i\pi}, \\
			 C_{2,1}\left(\tau\right)&\sim0.242\,\mathrm{e}^{i0.0242\pi}, \\
			 C_{3,1}\left(\tau\right)&\sim0.242\,\mathrm{e}^{-i0.0242\pi}.  
		\end{align*}
The value of $\left|C_{1,1}\right|$ reduces evidently under the DCE.
Similarly, we confirm whether Eq.~\eqref{eq:condition_A2} is satisfied for $m = 1$. 
		The left side of Eq.~\eqref{eq:condition_A2} becomes
		\begin{equation*}
			\left\{\,
			\begin{aligned}
				&\left|C_{2,1}\left(0\right)\right| - \left|C_{3,1}\left(0\right)\right| = 0 \\
				&\arg\left[C_{2,1}\left(0\right) + C_{3,1}\left(0\right) \right] - \arg \left[C_{1,1}\left(0\right)\right] = -\pi
			\end{aligned}
			\right.
		\end{equation*}
		and
		\begin{equation*}
			\left\{ \,
			\begin{aligned}
				&\left|C_{2,1}\left(\tau\right)\right| - \left|C_{3,1}\left(\tau\right)\right| = 0, \\
				&\arg\left[C_{2,1}\left(\tau\right) + C_{3,1}\left(\tau\right) \right] - \arg \left[C_{1,1}\left(\tau\right) \right]= -\pi.
			\end{aligned}
			\right.
		\end{equation*}
These results shows that $\arg\left[C_{2,1} + C_{3,1}\right] - \arg \left[C_{1,1}\right]$ does not change with time; 
		consequently, the condition of Eq.~\eqref{eq:condition_A2} is not satisfied.
		Hence, we conclude that synchronization in the balanced case is a consequence of the convergence of $C_{1,1}\left(t\right)$. 
		On the other hand, $\mathcal{B}_m \left(t\right) = \mathcal{C}_m \left(t\right)$ are always satisfied for $\Delta \theta = 0$ 
		since $C_{1,1}\left(0\right) = \frac{1}{2} \sin\left(\theta_1 - \theta_2\right) = 0$.
		Consequently, $\left| C_{1500}\left(500\right) \right|$ remains constant at 1 regardless of the values of $\alpha_0$ as shown in Fig.~\ref{fig:result1}(a).
		Furthermore, $\left| C_{1500}\left(500\right) \right|$ exhibits higher values in the regime of $\alpha_0 \le 4 \times 10^{-3}$ and $\Delta \theta /\theta_1 \sim 1$.
		In this regime, where $\theta_2 \sim 0$, anti-phase synchronization, which is unrelated to the DCE, appears~(see Appendix~\ref{sec:appendix_antisync}).

		In Figs.~\ref{fig:result1}(b)--(g), 
		we plot the time evolution of $\Braket{\sigma_\mathrm{z}^{1,2}}$, 
		$C_{\Delta t}\left(t\right)$, and the mean photon number $\Braket{n}\equiv\braket{a^\dagger a}$ 
		with $\theta_1 = \pi/4$, $\theta_2 = 11\pi/36$, and $\Delta t = 200$.
		We set $\alpha_0=0$ and $\alpha_0=6 \times 10^{-3}$ for Figs.~\ref{fig:result1}(b)--(d) and Figs.~\ref{fig:result1}(e)--(g), respectively.
		Comparing Fig.~\ref{fig:result1}(b) with Fig.~\ref{fig:result1}(e), 
		we can see that the degree of coincidence between $\Braket{\sigma_\mathrm{z}^{1}}$ and $\Braket{\sigma_\mathrm{z}^{2}}$
		is obviously higher in Fig.~\ref{fig:result1}(e) than in Fig.~\ref{fig:result1}(b).
		Correspondingly, the value of $C_{\Delta t}\left(t\right)$ approaches 1 in Fig.~\ref{fig:result1}(f).
		Moreover, Figs.~\ref{fig:result1}(b)--(c) show that the synchronization does not take place for $\alpha_0 = 0$, 
		despite the fact that the qubits interact with each other via the shared cavity. 
		This can be attributed to the presence of the dark state, as discussed in Sec.~\ref{sec:balanced}.
		
Shifting our focus to the growth of the mean photon number $\Braket{n}$ via the DCE, Figs.~\ref{fig:result1}(e)--(g)
		display an intriguing trend.
		The amplitudes of $\Braket{\sigma_\mathrm{z}^{1}}$ and $\Braket{\sigma_\mathrm{z}^{2}}$ decrease until $t=\tau$, 
		concomitant with the increase in the mean photon number. 
		As a consequence of the RWA, 
		we have neglected the terms on the time variation of the cavity resonance frequency, 
		namely $\alpha\left(t\right) a^{\dagger}a$, in Eq.~\eqref{eq:HI}.
		Thus, we confirm that DCE-induced photon generation, 
		rather than a change in the cavity resonant frequency, has an impact 
		on the dynamics of the qubits as well as the dynamics of the cavity itself, resulting in qubit synchronization. 

	\subsection{Unbalanced case with the same initial states} \label{sec:result2}
		\begin{figure}[t]
			\begin{minipage}[t]{1\hsize}
				\centering
				\includegraphics[width = 8.6cm]{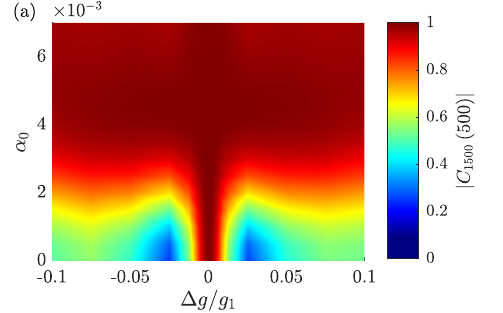}
			\end{minipage}\\
			\begin{minipage}[t]{1\hsize}
				\centering
				\includegraphics[width = 8.6cm]{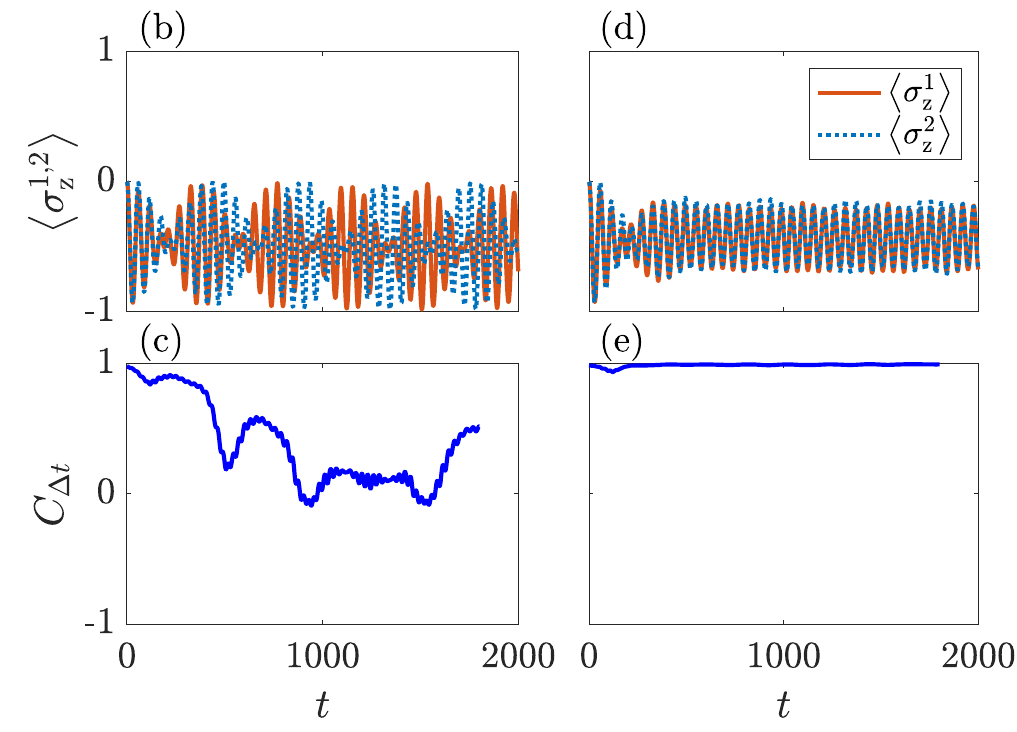}
			\end{minipage}
			\caption{Results of the unbalanced case with the same initial states.
				 We set $\theta_1 = \theta_2 = \pi/4$, $\omega = \omega_\mathrm{q}^1 = \omega_\mathrm{q}^2 = 1$, $\omega_\mathrm{d} = 2$, and $\tau = 500$.
				(a)~Pearson correlation coefficient $\left| C_{1500}\left(500\right) \right|$
				as a function of $\Delta g /g_1$ and $\alpha_0$, 
				where $\Delta g \equiv \left(g_1 - g_2\right)$ with $g_1 = 0.040$.
				(b) and (d) Time evolution of $\Braket{\sigma_\mathrm{z}^{1,2}}$. 
				(c) and (e) $C_{\Delta t}\left(t\right)$ with $\Delta t = 200$.
				We choose $\alpha_0=0$ for (b)--(c), and $\alpha_0=4 \times 10^{-3}$ for (d)--(e).
				For (b)--(e), the two qubits are coupled to the electromagnetic field with coupling strengths $g_1 = 0.040$ and $g_2 = 0.041$, respectively.
			}
			\label{fig:result2}
		\end{figure}
		
		Next, we show the synchronization between the qubits when their parameters are identical, 
		except for the coupling strengths $g_\mu$. 
		We set the values for $\theta_1$ and $\theta_2$ to $\pi/4$.
		Similar to Sec.~\ref{sec:result1}, the frequencies are set as $\omega = \omega_\mathrm{q}^1 = \omega_\mathrm{q}^2 = 1$ and $\omega_\mathrm{d} = 2$.
		In Fig.~\ref{fig:result2}(a), 
		we present the Pearson correlation coefficient $\left| C_{1500}\left(500\right) \right|$
		as a function of $\Delta g /g_1$ and $\alpha_0$, where $\Delta g \equiv \left(g_1 - g_2\right)$.
		The range of $\Delta g /g_1$ where $\left| C_{1500}\left(500\right) \right|$ exhibits higher values widens rapidly 
		around $\alpha_0 = 2\times10^{-3}$. 
		Furthermore, $\left| C_{1500}\left(500\right) \right|$ approaches 1 when $\alpha_0$ is approximately $4\times10^{-3}$. 
		However, around $\alpha_0 = 6\times10^{-3}$, $\left| C_{1500}\left(500\right) \right|$ is slightly lower than around $\alpha_0 = 4\times10^{-3}$.
		Put differently, synchronization cannot be enhanced by simply increasing $\alpha_0$.
		Dissimilar to the balanced case in Sec.~\ref{sec:result1}, 
		this results indicates that a moderate driving strength $\alpha_0$ and $\tau$ are required for strong synchronization.
		In the unbalanced case with the same initial states, 
		$C_1$, $C_2$, and $C_3$ in Eqs.~\eqref{eq:abcd_4a}--\eqref{eq:abcd_4c} become 
		\begin{align*}
C_1\left(0\right) &= \frac{g_2 - g_1}{G^2}\sin\theta\cos\theta, \\
C_2\left(0\right) &= C_3\left(0\right) = \frac{g_2 + g_1}{2 G^2}\sin\theta\cos\theta.
		\end{align*}
		If $\left|g_2-g_1\right| \ll g_{\mu}$, $\left|C_1\right| \ll \left|C_2\right|, \left|C_3\right|$ will hold.
		Thus, we conclude that $C_1$ is not the main factor inhibiting the synchronization.
		In this case, $C_{\mathcal{S},\mathcal{U}, 2}$ become crucial to inhibiting the synchronization.
		Calculating $C_{\mathcal{S},\mathcal{U}, 2}\left(\tau\right)$ 
		for $\alpha_0=4\times10^{-3}$ with $\theta_1=\theta_2=\pi/4$, $g_1 = 0.040$, and $g_2 = 0.041$, 
		we obtain 
		\begin{align*}
			\label{eq:C_result2}
			&C_{+1,+1, 2}\left(\tau\right)\sim0.0797\,\mathrm{e}^{i0.784\pi},\\
			&C_{+1,-1, 2}\left(\tau\right)\sim0.0843\,\mathrm{e}^{i0.916\pi}, \\
			&C_{-1,+1, 2}\left(\tau\right)\sim0.0979\,\mathrm{e}^{-i0.805\pi},\\
			&C_{-1,-1, 2}\left(\tau\right)\sim0.0827\,\mathrm{e}^{-i0.561\pi}.
		\end{align*}
		Clearly, $C_{-1, \mathcal{U}, m} = 0$, one of the two sufficient conditions for ensuring synchronization (see Sec.~\ref{sec:unbalanced}), 
		is not satisfied.
		On the other hand, the left side of Eq.~\eqref{eq:condition_B2}, 
		which corresponds to the other condition for synchronization, becomes 
		\begin{equation}
			\label{eq:C_result2_2}
			\left\{ \,
			\begin{aligned}
				& \left|C_{-1,+1, 2}\left(\tau\right)\right| - \left|C_{-1,-1, 2}\left(\tau\right)\right| \sim 0.0152 \\
				& \left|C_{+1,+1, 2}\left(\tau\right)\right| - \left|C_{+1,-1, 2}\left(\tau\right)\right| \sim -0.0046 \\
				& \arg\left[C_{-1, +1, 2} + C_{-1, -1, 2} \right] - \arg \left[C_{+1, +1, 2} + C_{+1, -1, 2} \right]\\
				&\qquad\qquad\qquad\qquad\qquad\qquad\quad \sim -1.55\pi.
			\end{aligned}
			\right.
		\end{equation}
		Although the slightly higher value of $\left|C_{-1,-1, 2}\right|- \left|C_{-1,+1, 2}\right|$ can be an obstacle to synchronization, 
		the condition ensuring $\Braket{\Delta\sigma_{\mathrm{z},m}\left(t\right)} \sim 0$ 
		shown in Eq.~\eqref{eq:condition_B2} is nearly satisfied. 
		This indicates that the synchronization of $\Braket{\sigma_\mathrm{z}^{1}}$ and $\Braket{\sigma_\mathrm{z}^{2}}$ at $\alpha_0 \sim 4 \times 10^{-3}$ 
		is caused by the rotation of $C_{\mathcal{S},\mathcal{U}, 2}\left(t\right)$ in the complex plane. 
		This rotational behavior seems to accounts for the reduction in $\left| C_{1500}\left(500\right) \right|$ for a regime of $\alpha_0 > 4 \times 10^{-3}$.

		Figures~\ref{fig:result2}(b)--(e)
		show examples of the time evolution of $\Braket{\sigma_\mathrm{z}^{1,2}}$ and 
		$C_{\Delta t}\left(t\right)$ with $g_1 = 0.040$, $g_2 = 0.041$, and $\Delta t = 200$.
		We set $\alpha_0=0$ and $\alpha_0=4 \times 10^{-3}$ for Figs.~\ref{fig:result2}(b)--(c) and Figs.~\ref{fig:result2}(d)--(e), respectively.
It is worth pointing out 
		that $\Braket{\sigma_\mathrm{z}^{1}}$ and $\Braket{\sigma_\mathrm{z}^{2}}$ in Fig.~\ref{fig:result2}(d) maintain in-phase synchronization 
		following the switching-off of the oscillation in the boundary condition at $t = 500 \left(=\tau\right)$.
		Correspondingly, the value of $C_{\Delta t}\left(t\right)$ maintains nearly 1 in Fig.~\ref{fig:result2}(e).
		This result may appear non-trivial because in the case shown in Figs.~\ref{fig:result2}(d)--(e),  
		each qubit has a distinct coupling strength $g_\mu$.
		Therefore, even if one qubit has the same initial state as the other, 
		they do not necessarily behave as a synchronized pair of oscillators. 
		Nonetheless, the stability of the synchronization can be ensured by satisfying the condition described by Eq.~\eqref{eq:condition_B2}. 
		As confirmed in Eq.~\eqref{eq:C_result2_2}, 
		this condition is almost satisfied by the parameters selected in Figs.~\ref{fig:result2}(d)--(e). 
		In essence, this underscores the fact that the stable synchronization depicted in Figs.~\ref{fig:result2}(d)--(e) is not an anomalous phenomenon; 
		rather, it represents a reasonable outcome.

	\subsection{Unbalanced case with different initial states} \label{sec:result3}
		\begin{figure}[t]
			\begin{minipage}[t]{1\hsize}
				\centering
				\includegraphics[width = 8.6cm]{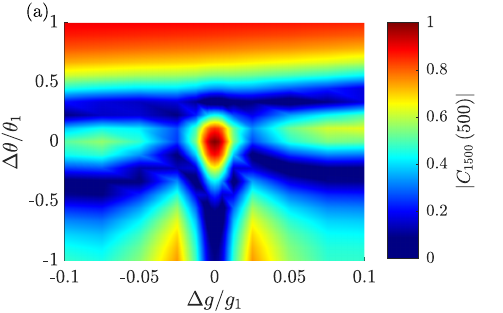}
			\end{minipage}\\
			\begin{minipage}[t]{1\hsize}
				\centering
				\includegraphics[width = 8.6cm]{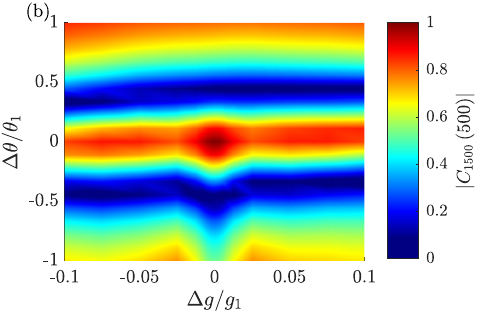}
			\end{minipage}\\
			\begin{minipage}[t]{1\hsize}
				\centering
				\includegraphics[width = 8.6cm]{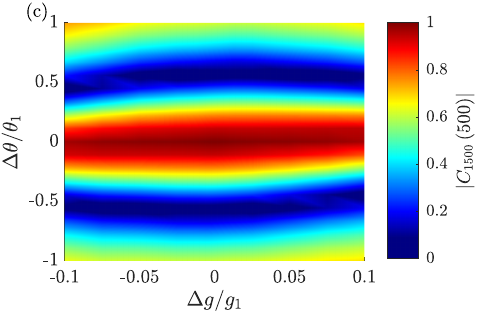}
			\end{minipage}
			\caption{Results of the unbalanced case with different initial states.
				We set $\omega = \omega_\mathrm{q}^1 = \omega_\mathrm{q}^2 = 1$, $\omega_\mathrm{d} = 2$, and $\tau = 500$.
				(a)--(c)~Pearson correlation coefficient $\left| C_{1500}\left(500\right) \right|$, 
				as a function of $\Delta g /g_1$ and $\Delta \theta / \theta_1$ 
				with (a)~$\alpha_0 = 0$, (b)~$\alpha_0 = 2.5 \times 10^{-3}$, and (c)~$\alpha_0 = 4.0 \times 10^{-3}$.}
				\label{fig:result3}
		\end{figure}
  
		Thus far, we have demonstrated synchronization under conditions where 
		either of $\theta_2$ or $g_2$ deviates from that of the other qubit.
		In this section, we explore the synchronization of qubits induced by the DCE under the scenario 
		where both $\theta_2$ and $g_2$ deviate from $\theta_1$ and $g_1$, respectively.
		Here, the Pearson correlation coefficient $\left| C_{1500}\left(500\right) \right|$ as a function of $\Delta g /g_1$ and $\Delta \theta / \theta_1$ is shown in Fig.~\ref{fig:result3}.
The driving strengths are set as $\alpha_0 = 0$ in Fig.~\ref{fig:result3}(a), $\alpha_0 = 2.5 \times 10^{-3}$ in Fig.~\ref{fig:result3}(b), 
		and $\alpha_0 = 4.0 \times 10^{-3}$ in Fig.~\ref{fig:result3}(c). 
		We set $\theta_1 = \pi/4$, $g_1 = 0.04$ and deviate $\theta_2$ and $g_2$ from $\theta_1$ and $g_1$, respectively. 
		We observe that the synchronization in the regime of $\theta_1 \sim \theta_2$ 
		is stronger in Fig.~\ref{fig:result3}(b) than in Fig.~\ref{fig:result3}(a).
		Similarly it is stronger in Fig.~\ref{fig:result3}(c) than in Fig.~\ref{fig:result3}(b).
		
		It should be noted here that the region 
		where $\left| C_{1500}\left(500\right) \right|$ shows greater values
		expand almost horizontally in Figs.~\ref{fig:result3}(b)--(c).
		This phenomenon is arguably unique to DCE driving; 
		it does not occur when microwave driving is employed~(see Appendix~\ref{sec:appendix_linear}). As discussed in Sec.~\ref{sec:result1} and Sec.~\ref{sec:result2}, 
$C_{j,1}$ in Eqs.~\eqref{eq:abcd_2a}--\eqref{eq:abcd_2c} 
		significantly impact the occurrence of synchronization in the balanced scenario with the different initial states, 
		whereas $C_{\mathcal{S},\mathcal{U},2}$ in Eqs.~\eqref{eq:abcd_3a}--\eqref{eq:abcd_3d} significantly impact its occurrence in unbalanced scenarios 
		given that $\left|\Delta g\right| \ll g_1, g_2$.
		In other words, $\mathcal{B}_0$ and $\mathcal{C}_0$ primarily affect the synchronization when $\theta_1 \ne \theta_2$, 
		whereas $\mathcal{B}_1$ and $\mathcal{C}_1$ primarily affect it when $g_1 \ne g_2$.
		Here, we notice that 
		no coupling exist between $\left\{\mathcal{B}_0, \mathcal{C}_0\right\}$ and $\left\{\mathcal{B}_1, \mathcal{C}_1\right\}$ for $t > \tau$ 
		because the Schr\"{o}dinger equation in the rotating frame forms independent systems of differential equations
		within $\mathcal{A}_{m+1}$, $\mathcal{B}_m$, $\mathcal{C}_m$, and $\mathcal{D}_{m-1}$.
		Meanwhile, for $t \le \tau$, the external driving induces coupling 
		between $\left\{\mathcal{A}_{m+1},\mathcal{B}_m,\mathcal{C}_m,\mathcal{D}_{m-1}\right\}$; 
however, it is important to note that the driving term is expressed as $\frac{\alpha_0}{2} ({a^{\dagger}}^2 + a^2)$, 
		indicating that $\left\{\mathcal{A}_{m+1},\mathcal{B}_m,\mathcal{C}_m,\mathcal{D}_{m-1}\right\}$ are coupled alternately. Specifically, the Schr\"{o}dinger equation for $t \le \tau$ in a frame rotating at the driving frequency $\omega_\mathrm{d}$ 
		is disentangled into two sets of differential equations: 
		one for $\left\{\mathcal{A}_{j+1},\mathcal{B}_j,\mathcal{C}_j,\mathcal{D}_{j-1}\right\}$, 
		where $j$ are odd numbers, 
		and the other for $\left\{\mathcal{A}_{l+1},\mathcal{B}_l,\mathcal{C}_l,\mathcal{D}_{l-1}\right\}$, 
		where $l$ are even numbers. 
		Consequently, $\left\{\mathcal{B}_0,\mathcal{C}_0\right\}$ and $\left\{\mathcal{B}_1,\mathcal{C}_1\right\}$ are governed by separate, 
		independent sets of differential equations for $t > \tau$ as well as $t \le \tau$ .
		For this reason, the differences in $\theta_{\mu}$ and $g_{\mu}$ independently affect the synchronization 
		without any overlap between these factors.
		Consequently, the region with higher values of $\left| C_{1500}\left(500\right) \right|$ expands horizontally.
		
	\subsection{Balanced case with the same initial states and frequency detuning} \label{sec:result4}
		\begin{figure}[t]
			\begin{minipage}[t]{1\hsize}
				\centering
				\includegraphics[width = 8.6cm]{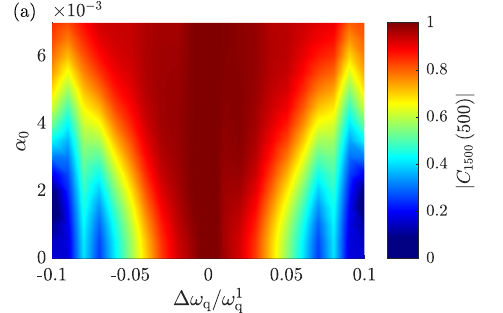}
			\end{minipage}\\
			\begin{minipage}[t]{1\hsize}
				\centering
				\includegraphics[width = 8.6cm]{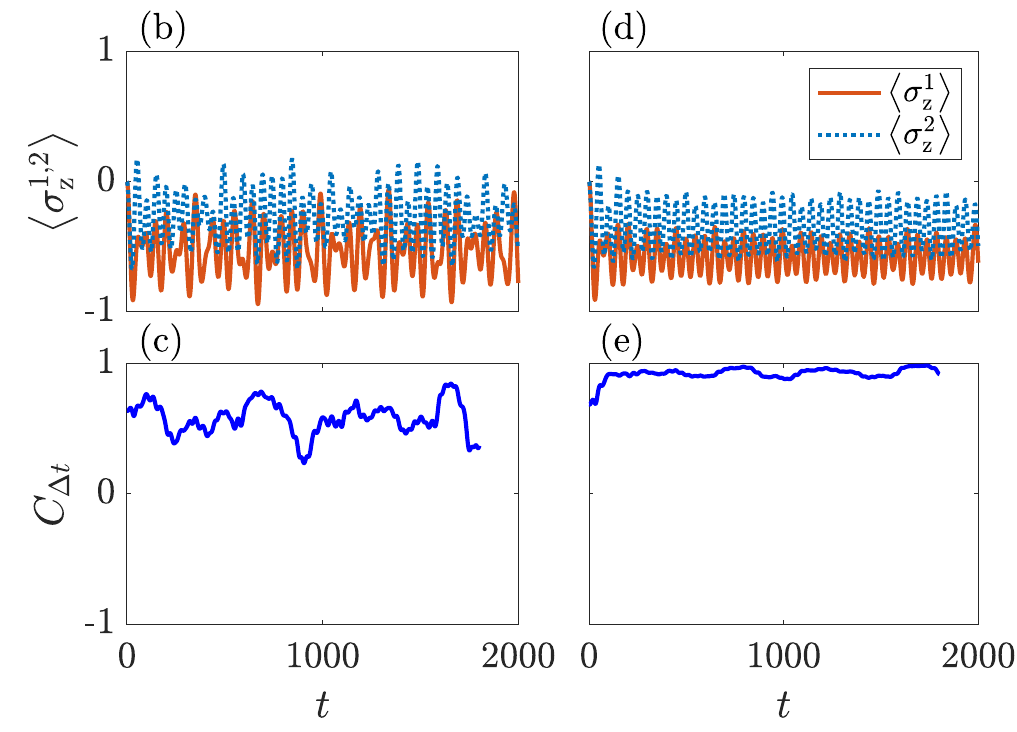}
			\end{minipage}
			\caption{Results of the balanced case with the same initial states and frequency detuning.
				We set $\theta_1 = \theta_2 = \pi/4$, $g_1 = g_2 = 0.04$, and $\tau = 500$.
				(a)~Pearson correlation coefficient $\left| C_{1500}\left(500\right) \right|$
				as a function of $\Delta \omega_\mathrm{q} / \omega_\mathrm{q}^1$ and $\alpha_0$. 
				where $\Delta \omega_\mathrm{q} \equiv \left(\omega_\mathrm{q}^1 - \omega_\mathrm{q}^2\right)$ with $\omega_\mathrm{q}^1 = 1$. 
				(b) and (d) Time evolution of $\Braket{\sigma_\mathrm{z}^{1,2}}$. 
				(c) and (e) $C_{\Delta t}\left(t\right)$ with $\Delta t = 200$.
				We select $\alpha_0=0$ and $\alpha_0=6 \times 10^{-3}$ for (b)--(c) and (d)--(e), respectively.
				For (b)--(e), We set $\omega_\mathrm{q}^1 = 1.00$ and $\omega_\mathrm{q}^2 = 1.05$.
			}
			\label{fig:result4}
		\end{figure}
	
		Finally, we present the numerical results illustrating qubit synchronization in the presence of the frequency detuning 
		with $\theta_1 = \theta_2 = \pi/4$ and $g_1 = g_2 = 0.04$.
		In this context, the exact solutions for $\mathcal{A}_m, \mathcal{B}_m, \mathcal{C}_m$, and $\mathcal{D}_m$
		are not as clear as those presented in Sec.~\ref{sec:analitical}. Therefore, we only consider the numerical results.
		In Fig.~\ref{fig:result4}(a), 
		we plot the Pearson correlation coefficient $\left| C_{1500}\left(500\right) \right|$
		as a function of $\Delta \omega_\mathrm{q} /\omega_{\mathrm{q}}^1$ and $\alpha_0$, 
		where $\Delta \omega_\mathrm{q} \equiv \left(\omega_\mathrm{q}^1 - \omega_\mathrm{q}^2\right)$.
		The range of $\Delta \omega_\mathrm{q}$ 
with higher values of $\left| C_{1500}\left(500\right) \right|$ widens gradually as $\alpha_0$ increases.
		Apparently, the rapid expansion of $\left| C_{1500}\left(500\right) \right|$, 
		which occurs in the unbalanced case with different initial states~(see Fig.~\ref{fig:result2}), 
		is not in this scenario.
		This result indicates that the greater the driving strength, 
		the stronger the correlation of the qubits, at least within the region of $\alpha_0 \le 7 \times 10^{-3}$.

		In Figs.~\ref{fig:result4}(b)--(e),
		we plot the time evolution of $\Braket{\sigma_\mathrm{z}^{1,2}}$ and
		$C_{\Delta t}\left(t\right)$ with $\omega_{\mathrm{q}}^1=1.00$, $\omega_{\mathrm{q}}^2=1.05$, and $\Delta t = 200$.
		We set $\alpha_0=0$ and $\alpha_0 = 6 \times 10^{-3}$ for Figs.~\ref{fig:result4}(b)--(c) and Figs.~\ref{fig:result4}(d)--(e), respectively.
Despite the persistence of an offset between $\Braket{\sigma_\mathrm{z}^{1}}$ and $\Braket{\sigma_\mathrm{z}^{2}}$, following driving by the DCE, 
they
		synchronize evidently even for $t > 500$, as shown in Figs.~\ref{fig:result4}(d)--(e).
This result suggests the existence and satisfaction of the condition for ensuring stable synchronization,
		similar to other cases.

\section{discussion} \label{sec:discussion}
	In Sec.~\ref{sec:analitical}, 
	we have theoretically analyzed the behavior of the two-atom TCM. 
	In the absence of the DCE, 
	namely in $t > \tau$, 
we have derived analytical solutions for the time evolution of the system 
	assuming the two qubits and cavity are in resonance 
	(Eqs.~\eqref{eq:abcd_1a}--\eqref{eq:abcd_2c} and Eqs.~\eqref{eq:abcd_3a}--\eqref{eq:abcd_4c}). 
	Thereafter, we have derived the sufficient conditions for achieving synchronization in balanced and unbalanced scenarios without frequency detuning, 
regarding coefficients $C_{j,m}$, $C_{\mathcal{S},\mathcal{U}, m}$, 
	and $C_{j}$ in Eqs.~\eqref{eq:abcd_1a}--\eqref{eq:abcd_2c} and Eqs.~\eqref{eq:abcd_3a}--\eqref{eq:abcd_4c}. 
	These coefficients are time-dependent for $t \le \tau$ in the presence of the DCE. 
	To fulfill the conditions for in-phase synchronization, at least one of the following is required: 
	\begin{enumerate}
		\item Convergence of specific coefficients to 0. \label{item:cond1}
		\item Moderate rotation of the coefficients in the complex plane. \label{item:cond2}
	\end{enumerate}
	In Sec.~\ref{sec:numerical_results}, we have calculated and analyzed the evolution 
of the system.
	Synchronization occurs both in the balanced and unbalanced cases without frequency detuning. 
	In the balanced case without frequency detuning, 
	we have observed convergence of a specific coefficient. 
	Consequently, the Pearson correlation coefficient $\left| C_{1500}\left(500\right) \right|$, 
	a metric of synchronization, 
	increases monotonically with the driving strength, as illustrated in Fig.~\ref{fig:result1}.
	However, in the unbalanced case, we have witnessed a counterintuitive phenomenon: 
	$\left| C_{1500}\left(500\right) \right|$ does not increase monotonically with the driving strength, 
	as depicted in Fig.~\ref{fig:result2}. 
	This result indicates that the moderate rotations of the coefficients 
	in the complex plane are crucial to the occurrence of synchronization when $g_1 \ne g_2$.
	
	In Sec.~\ref{sec:result1}, where we presented the numerical results of the balanced case with different initial states, 
	we have observed that the photon generation in the cavity caused by the DCE has affected the behavior of the qubits, 
	resulting in their synchronization. 
	Moreover, in Sec.~\ref{sec:result2}, we have analyzed the qubit synchronization in the unbalanced case with the same initial states, 
	confirming the occurrence of stable synchronization, 
	as predicted by the theoretical analysis in Sec.~\ref{sec:unbalanced}. 
	
	Furthermore, our investigation in Sec.~\ref{sec:result3} have revealed a remarkable phenomenon: 
	the differences in the $\theta_{\mu}$ and $g_\mu$ independently influence the synchronizationwithout the overlapping of both factors.
	We have concluded that this intriguing phenomenon
is a unique feature of DCE-induced synchronization
	stemming from the squared terms of the creation and annihilation operators expressing the DCE in Eq.~\eqref{eq:H}.
	
	In Ref.~\cite{Zhirov2}, the authors investigated the synchronization of two qubits coupled to a driven dissipative resonator.
Although their experimental setup has similarities with ours, 
	they primarily analyzed the adjustment of the frequency and phase between the external drive and $\Braket{\sigma_\mathrm{x}}$ of individual qubits
in the balanced scenario, where $g_1=g_2$.
In contrast, our research has delved into the synchronization of $\Braket{\sigma_\mathrm{z}}$ of two qubits 
	in balanced and unbalanced scenarios, 
	exploring additional aspects that go beyond the scope of their work. 
Regarding the TCM not coupled to an external bath, 
	the study by Ref.~\cite{huan2020synchronization} is closely related to ours. 
	However, we have discussed the TCM in more detail and examined the conditions for ensuring synchronization. 
	Moreover, our study distinguishes itself from the previous studies, 
	as it has specifically addressed the phenomena associated with the DCE.
	Consequently, our outcomes represent significant progress over the extant studies.

\section{Conclusion} \label{sec:conclusion}
	Summarizing, we have investigated the quantum synchronization of two qubits induced by the DCE, 
	employing the two-atom TCM as our framework.
	First, we have theoretically analyzed the behavior of the two-atom TCM, 
	revealing the sufficient conditions to ensure synchronization in balanced and unbalanced scenarios. 

	Afterward, we have presented the numerical results.
	In the balanced scenario, 
	where the coupling strengths are uniform but the initial state of one qubit diverges from the other ($g_1=g_2$, $\theta_1\ne\theta_2$), 
	we have demonstrated that DCE-induced photon generation in the cavity has an impact on the qubit behaviors, 
resulting in qubit synchronization. 
	In the unbalanced scenario, 
	where each qubit interacts with the cavity through different coupling strengths ($g_1\ne g_2$), 
	the synchronization persists even after turning off the DCE, 
	as predicted by our theoretical analysis of the TCM. 
	Notably, our findings have unveiled
	a remarkable characteristic of DCE driving: 
	the differences in $\theta_{\mu}$ and $g_\mu$ independently affect qubit synchronization without overlapping.
	Furthermore, we have demonstrated the synchronization in the presence of the frequency detuning.
	In this situation, we have observed an offset between 
	$\Braket{\sigma_\mathrm{z}^{1}}$ and $\Braket{\sigma_\mathrm{z}^{2}}$, 
	which differs from the results obtained under other conditions. 
	Despite the persistence of the offset after DCE driving, the DCE induces evident qubit synchronization.

	Although significant insights have been gained from this study, 
	some aspects, such as the underlying mechanism accounting for the obtained synchronization behavior
	as well as the potential applications of this phenomenon in quantum information processing and quantum communication, 
	is still obscure. 
	These would form the basis of our future study.

\begin{acknowledgments}
	The fruitful comments by Wataru Setoyama are greatly appreciated. 
	This work was supported by JSPS KAKENHI Grant Number JP22H03659.
\end{acknowledgments}

\appendix
\section{Anti-phase synchronization irrelevant to the DCE} \label{sec:appendix_antisync}
		\begin{figure}[t]\centering
			\includegraphics[width = 8.6cm]{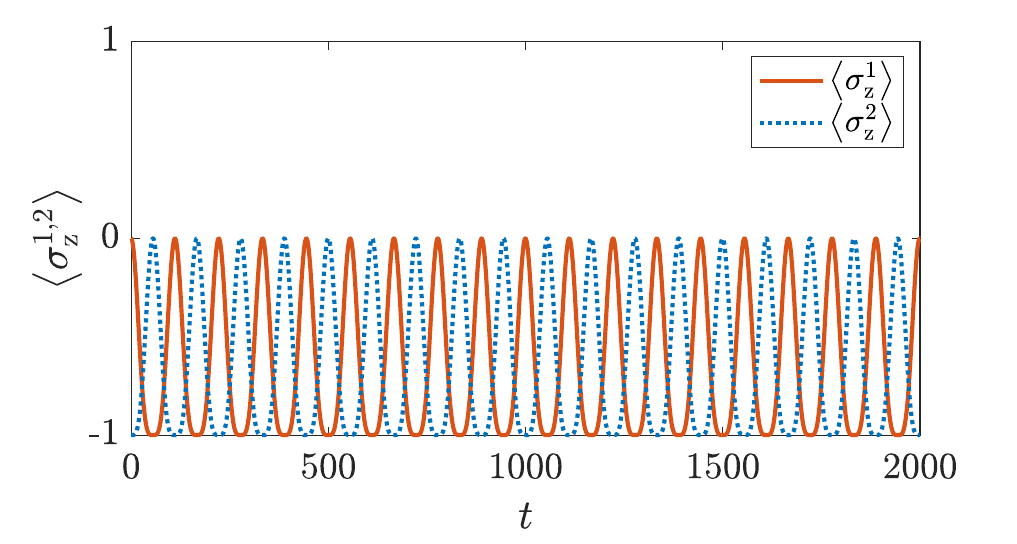}
			\caption{Time evolution of $\Braket{\sigma_\mathrm{z}^{1,2}}$ in the balanced case where $g_1=g_2$. 
				We set $\alpha_0=0$, $\theta_1 = \pi/4$, $\theta_2 = 0$, $g_1=g_2=0.04$, 
				$\omega = \omega_\mathrm{q}^1 = \omega_\mathrm{q}^2 = 1$, and $\omega_\mathrm{d} = 2$.
				Anti-phase synchronization observed in this figure is irrelevant to the DCE.}
			\label{fig:pauli_z_initialTheta}
		\end{figure}
		In the balanced scenario (Sec.~\ref{sec:result1}), where $g_1=g_2$, 
		$\left| C_{1500}\left(500\right) \right|$ exhibits a higher value in the regime of $\alpha_0 \le 4 \times 10^{-3}$ and $\theta_2 \sim 0$,
		as shown in Fig.~\ref{fig:result1}(a).
		Figure~\ref{fig:pauli_z_initialTheta} shows the time evolution of $\Braket{\sigma_\mathrm{z}^{1,2}}$ 
		for $\alpha_0=0$, $\theta_1 = \pi/4$, $\theta_2 = 0$, and $g_1=g_2=0.04$.
At $t = 0$, 
		$\Braket{\sigma_\mathrm{z}^{2}}$ begins transitioning to the excited state from the outset,
		whereas $\Braket{\sigma_\mathrm{z}^{1}}$ begins transitioning to the ground state.
		Therefore, in the balanced scenario, two qubits exchange energy through the cavity in a constant rhythm akin to playing catch, resulting in anti-phase synchronization.
		This phenomenon is irrelevant to the DCE and beyond the scope of this paper; 
		thus, it will not be explored in detail.

	\section{Synchronization induced by microwave driving} \label{sec:appendix_linear}
		\begin{figure}[t]
			\begin{minipage}[t]{1\hsize}
				\centering
				\includegraphics[width = 8.6cm]{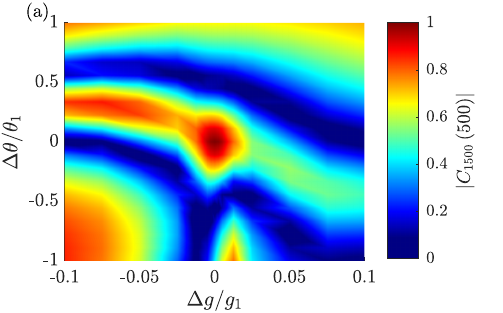}
			\end{minipage}\\
			\begin{minipage}[t]{1\hsize}
				\centering
				\includegraphics[width = 8.6cm]{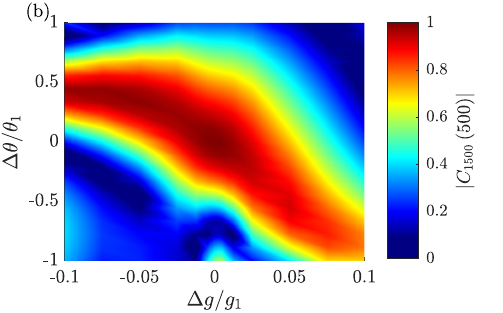}
			\end{minipage}\\
			\caption{Results of synchronization induced by microwave driving in the unbalanced case with different initial states.
					The system Hamiltonian is expressed as Eq.~\eqref{eq:H_Appendix}.
					We set $\omega = \omega_\mathrm{q}^1 = \omega_\mathrm{q}^2 = \omega_\mathrm{d} = 1$, and $\tau = 500$.
					(a)--(b)~Pearson correlation coefficient $\left| C_{1500}\left(500\right) \right|$
					as a function of $\Delta g /g_1$ and $\Delta \theta / \theta_1$ 
					with (a)~$\alpha_0 = 4 \times 10^{-3}$ and (b)~$\alpha_0 = 7 \times 10^{-3}$. When $\alpha_0 = 0$, 
					$\left| C_{1500}\left(500\right) \right|$ displays exactly the same behavior as that in Fig.~\ref{fig:result3}(a); 
					therefore, it is omitted here.}
			\label{fig:arnold_lineardrive}
		\end{figure}
In Sec.~\ref{sec:result3}, we have discussed the unique feature of DCE-induced synchronization, 
		that is, the differences in $\theta_{\mu}$ and $g_\mu$ independently affect qubit synchronization without overlapping. 
		Because of this feature, the region with higher values of $\left| C_{1500}\left(500\right) \right|$ expands horizontally in Fig.~\ref{fig:result3}(c).
		We have concluded that this intriguing feature results from the squared terms of the creation and annihilation operators 
		in Eq.~\eqref{eq:H}, which represents DCE driving.
		To substantiate this conclusion, 
		we briefly examine synchronization induced by microwave driving.

		Here, we define the Hamiltonian describing the TCM with the microwave driving as follows~\cite{blais2004cavity}:
		\begin{equation}
			\label{eq:H_Appendix}
			H_{\mathrm{w}} \equiv H_{\mathrm{TC}} + \alpha\left(t\right) \left( a^{\dagger}+a \right),
		\end{equation}
		where $\alpha \left(t\right) = \alpha_0 \cos\left(\omega_\mathrm{d}t\right) \Theta\left( \tau - t \right)$. 
In the frame rotating at the driving frequency $\omega_{\mathrm{d}}$, 
		the fast oscillating terms can be neglected using the RWA, 
		allowing the system Hamiltonian to be written as
		\begin{align*}
			\label{eq:HI_Appendix}
			H_{\mathrm{J}} \equiv
			\left(\omega - \omega_{\mathrm{d}}\right)a^{\dagger}a
			&+\sum_{\mu=1}^2{ \left[\frac{\omega_{\mathrm{q}}^{\mu} - \omega_{\mathrm{d}}}{2}
			+ g_{\mu} \left( \sigma_{\mathrm{-}}^\mu a^{\dagger} + \sigma_{\mathrm{+}}^\mu a \right)\right]} \nonumber\\
			&+ \frac{\alpha_0}{2} \Theta\left( \tau - t \right) \left( {a^{\dagger}} + a\right).
		\end{align*}  
		Thereafter, the time evolution of the system is governed by the Schr\"{o}dinger equation, i.e., 
		$\dot{\Ket{\psi_{\mathrm{I}}}} = -i H_{\mathrm{J}} \Ket{\psi_{\mathrm{I}}}$.

		In Fig.~\ref{fig:arnold_lineardrive}, 
		we plot $\left| C_{1500}\left(500\right) \right|$ as a function of $\Delta g /g_1$ and $\Delta \theta / \theta_1$ 
		in the absence of frequency detuning ($\omega_{\mathrm{q}}^1 = \omega_{\mathrm{q}}^2 = \omega = {\omega_{\mathrm{d}}} = 1$).
We fix $\alpha_0 = 4 \times 10^{-3}$ and $\alpha_0 = 7 \times 10^{-3}$ 
		for Fig.~\ref{fig:arnold_lineardrive}(a) and Fig.~\ref{fig:arnold_lineardrive}(b), respectively.
A comparison of Fig.~\ref{fig:arnold_lineardrive} and Fig.~\ref{fig:result3} reveals a notable difference.
		In Fig.~\ref{fig:arnold_lineardrive}, the region displaying stronger synchronization exhibits evident bends.
		This is because 
		the linear terms of the creation and annihilation operators in the driving term of Eq.~\eqref{eq:H_Appendix} 
		couple $\left\{\mathcal{B}_0,\mathcal{C}_0\right\}$ and $\left\{\mathcal{B}_1,\mathcal{C}_1\right\}$.
		Further discussion of the relation between synchronization and $\left\{\mathcal{B}_m,\mathcal{C}_m\right\}$ is presented in Sec.~\ref{sec:result3}.
		In conclusion, 
		the characteristic that the differences in $\theta_{\mu}$ and $g_\mu$ independently influence synchronization 
does indeed arise from DCE driving, but not from microwave driving.

\bibliography{DCE, classical_sync, quantum_sync, qubit, extra}

\end{document}